# Proposal of a spin torque majority gate logic

Dmitri E. Nikonov, George I. Bourianoff, and Tahir Ghani

Components Research, Intel Corp., 2200 Mission College Blvd., Santa Clara, California 95052, USA

Components Research, Intel Corp., 5000 Plaza on the Lakes Blvd., Austin, Texas 78746, USA

Portland Technology Development, Intel Corp., 2501 NW 229th Ave., Hillsboro, Oregon 97124, USA

A new spin based logic device is proposed. It is comprised of a common free ferromagnetic layer separated by a tunnel junction from three inputs and one output with separate fixed layers. It has the functionality of a majority gate and is switched by spin transfer torque. Validity of its logic operation is demonstrated by micromagnetic simulation. A version of such devices with perpendicular magnetization is examined. Switching encompasses moving domain walls. The device reuses most of the materials and structures from spin torque RAM, and is entirely compatible with CMOS technology.

Spin based devices are one of the alternative computing technologies listed in the International Technology Roadmap for Semiconductors[1]. Research in spintronics[2] has resulted in fascinating fundamental physics discoveries and in proposals for several spintronic devices[3]. Some of the devices belong to the class with electric inputs and outputs and a spin degree of freedom involved in their operation: Datta-Das spin modulator[4], spin FET[5], all-spin-logic (ASL) device[6], magnetic tunnel junction (MTJ) based logic[7,8,9,10]. Devices of this class are completely defined by their electrical inputs and outputs. The other class consists of proper spintronic devices defined by spin inputs and outputs e.g. magnetic cellular automata (MQCA)[11], magnetic domain wall logic[12], domain wall majority gates (DWMG)[13], spin wave bus (SWB) devices[14,15], spin gain transistor[16]. In this second class of devices it is necessary to convert spin to electrical output and vice versa, in order to communicate with the electronic circuits. However within the spintronic circuit the computational variable is stored and the signal is passed from one device to another in the spin form.

A new type of spin logic device – spin torque majority gate (STMG) is proposed in this paper. Arguments are provided showing that STMG has advantages compared to previously proposed spin-based logic. It has three inputs and one output. The output of the majority gate assumes the same logical state ("0" or "1") as the majority of the three inputs. For the principle of operation and a truth table of a majority gate see the example of quantum cellular automata (QCA)[17]. Magnetization of a common free layer is switched by spin torque to a state determined by the majority (at least 2 of the 3) of current passed into its inputs. The stack of layers in STMG with in plane magnetization is similar to that in a MTJ, though the inputs are represented by separate nanopillars electrically isolated from each other (Figure 1a). The structure bears some resemblance to three terminal spin transfer torque random access memory (STTRAM)[18], and is compatible with CMOS process[19]. However STMG has 5 electrodes and does not include a spin valve. Moreover, unlike STTRAM, STMG can implement complicated logic functions. This device has a common free layer that is switched upon application of sufficient spin torque. Each of the input nanopillars and the output nanopillar has separate fixed layers which are pinned by the anti-ferromagnetic layers on top of them. The one significant difference in processing compared to STTRAM is the etching of nanopillars. The top view of the device is shown in Figure 1b. The free layer is shaped as an ellipse in order to create two stable states (logical "0"and "1") along the plus and minus directions of the long axis due to shape anisotropy. To maintain the approximate symmetry between the input electrodes, they are placed at the periphery of the free layer and the output electrode is placed in the center of STMG.

STMG is different from MTJ-based-logic where MTJ is used to just stores a bit, but logic is done in transistors. In STMG, the logic function is performed in the common free magnetic layer. The three input currents are not combined, but rather act at different locations of the free layer. STMG is also different from MQCA and DWMG in that it does not require the external clocking magnetic field. This is an important advantage both from the energy consumption and the packaging design standpoint. Even though ASL use spin torque as well, it plays a different role there – to maintain a nanomagnet in an unstable state until a signal arrives to switch it. In ASL the signal is passed as spin-polarized current, which is known to be fraught with challenges[20]. In STMG, the signal is passed as a wave of magnetization. Perpendicular magnetization allows one to decrease the critical current of the spin torque devices[21], and therefore to lower power dissipation, which makes magnetic circuits more competitive. STMG with perpendicular magnetization is also proposed. Switching in it occurs due to motion of domain walls. In this sense it is somewhat reminiscent to domain wall memory[22], though in STMG current passes perpendicular to ferromagnetic wires, rather than along wires. It is also reminiscent of the domain wall logic[12], though STMG does not require an alternating external magnetic field and has different logic functionality.

The polarity of the voltage applied to each of the input nanopillars (Figure 1b) corresponds to the logical state of the input. If the input voltage to the nanopillar is negative, the current goes from bottom to top, and it forces the magnetization in the free layer below the nanopillar to align parallel with the magnetization of the fixed layer which is uniform for all the nanopillars. If the voltage is positive, the current forces the magnetization of the nanopillar to align opposite to the fixed layer. These local influences interact with each other and set the alignment of the total magnetization to the alignment of the majority of nanopillars, which provides the desired gate logic functionality. In an integrated circuit, a driving transistor might be needed to provide current to each nanopillar. Once the direction of magnetization under the output nanopillar settles close to its steady state value, the input current can be turned off, and the direction of magnetization can be determined from the magnetoresistance of the stack underneath the output nanopillar. The output signal can be detected using tunneling magnetoresistance effect where parallel alignment of the free and fixed layer corresponds to a lower resistance, while anti-parallel alignment corresponds to a higher resistance. This detection mechanism is combined with an output sense amplifier which is borrowed from the well-developed STTRAM technology. Note that the final state does not depend on the initial state, but only on the directions of the input currents. Therefore there is no need to read the state of the STMG before switching, which is a very useful attribute.

Our simulation is based on the Landau-Lifshitz-Gilbert equation[23] for magnetization. Effective magnetic field comes from the gradient of magnetic energy comprised of the demagnetization energy, material anisotropy, and exchange energy[23]. Simulations are performed using the NIST simulator OOMMF[24]. These simulations are used to verify that the device indeed performs switching with the desired logic functionality. An example of such a simulation with polarities (A+,B+,C-) is presented in Figure 2, showing the distribution of magnetization at various times after the start of the input current pulses. The local direction of magnetization is highlighted both by the direction of the arrows and the color (right – red, left – light blue, up – yellow, bottom- dark blue). The out of plane magnetization component is small and is not shown here. The size of the STMG is 120x90x3nm, magnetization $M_s = 1.42 \cdot 10^6 A/m$ corresponding to Co, Gilbert damping $\alpha = 0.014$, current $I = 4mA$ through each nanopillar, polarization $P = 0.57$, and exchange constant $A = 2 \cdot 10^{-11} J/m$. Magnetization switches in a very complicated pattern: it starts near the two right nanopillars, propagates to the left and bounces several times around the ellipse. Then it crosses over to the left direction and settles close to it after a few ringing cycles. Since the two right pillars promote 180 degree turning of magnetization, they win the fight with one left pillar, and switching of magnetization to the left direction happens as expected. Note that the distribution of magnetization is non-uniform and cannot be captured by a simpler macrospin model. Simulations with all other combinations of voltage polarities have verified that indeed the STMG works in accordance with the truth table.

A similar STMG device can also be fabricated with perpendicular magnetization of the ferromagnetic layers. Its cross-section is presented in Figure 3a. Materials with perpendicular magnetization have a significant material anisotropy which makes the state with out of plane magnetization have less energy. When the material anisotropy exceeds the shape anisotropy, magnetizations tends to point out of plane. Here the stable equilibrium states corresponding to logical "0" and "1" are the magnetization directions up and down, i.e., out of plane of the chip.

The areas with magnetization pointing up are separated from those pointing down by clearly defined domain walls. One can also separate the areas of inputs from the output and place the output on the periphery of the device, arriving at the cross configuration, structure shown in Figure 3b.

An example of simulation of switching of a cross STMG is shown in Figure 4. A different color scheme is used to denote out of plane magnetization for this figure: but arrows still designate the direction of the in plane projections of magnetization. The color corresponds to the projection on the vertical axis: red – up, white – zero, blue – down. The width of the wires is 30nm, so the span of the cross is 150x150x3nm. The parameters are: $M_s = 4 \cdot 10^5 A/m$, $\alpha = 0.007$, $I = 0.5mA$, $P = 0.9$, $A = 2 \cdot 10^{-11} J/m$, material anisotropy $K_u = 91 kJ/m^3$. It demonstrates how the spin torques at electrodes moves the domain walls along the wires. It has been reported that current perpendicular to the plane is more efficient in moving domain walls than current along the ferromagnetic wires[25]. This illustrates a different nature of switching in cross STMG: the input spin torques, aiming to flip magnetization, first win over their respective arms of the cross. Torques with the opposite polarity aim to preserve the initial magnetization direction in their respective arm. The net result is that the majority of the input torques win over the middle of the cross (either flipping or preserving magnetization there). Then the domain wall propagates to the output arm of the cross and the majority of inputs enforce their magnetization direction at the output. When the current is switched off, the opposing input is conquered by the majority.

In conclusion, a novel spin logic device – spin torque majority gate is proposed, its logic operation is verified by micromagnetic simulation. A cross STMG with perpendicular polarization is shown to have identical logic functionality. STMG has advantages over other spin-based logic devices in simplicity of operation and CMOS compatibility.

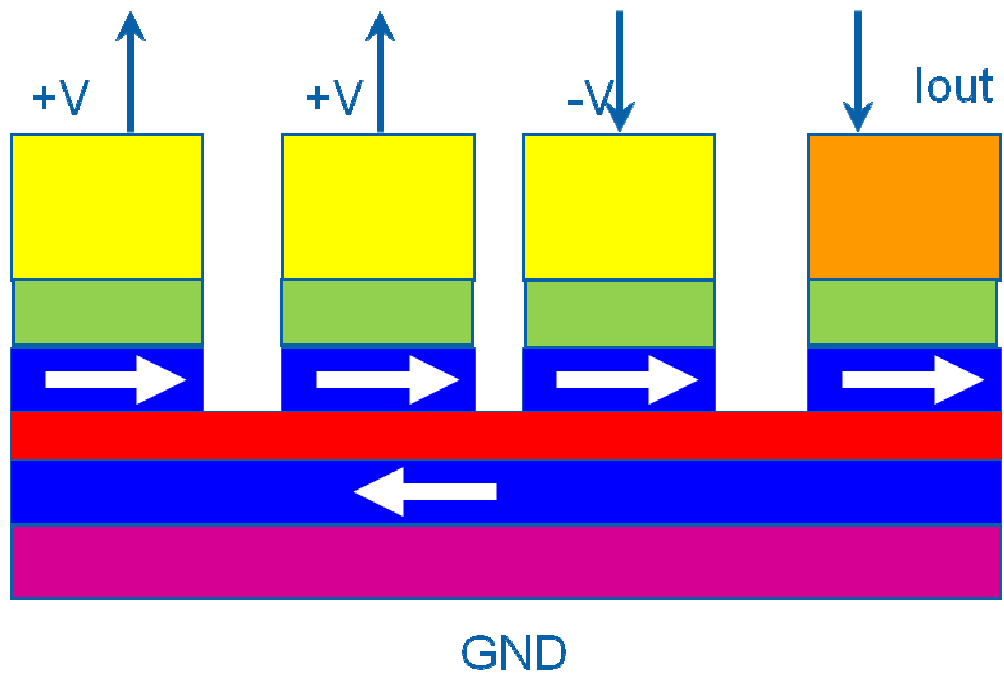

Figure 1a. The schematic cross-section of STMG. The lower layer with a white arrow (magnetization) is the free FM layer. The set of higher layers with arrows are the fixed FM layers for inputs (designated by applied voltages) and the output (designated by the output current). Three input nanopillars (on the left) are designated by input voltages, and one output nanopillar (on the right) is designated by the output current.

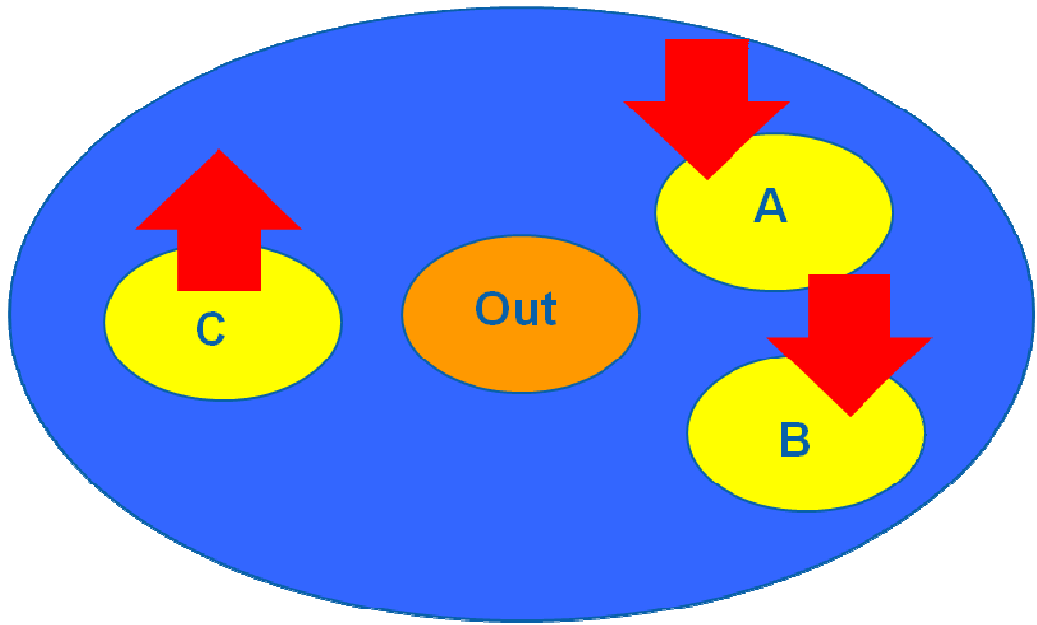

Figure 1b. The schematic top view of STMG. Outer ellipse is the common free layer. Ellipses marked with "A", "B", and "C" are input nanopillars. The ellipse marked with "Out" is the output nanopillars.

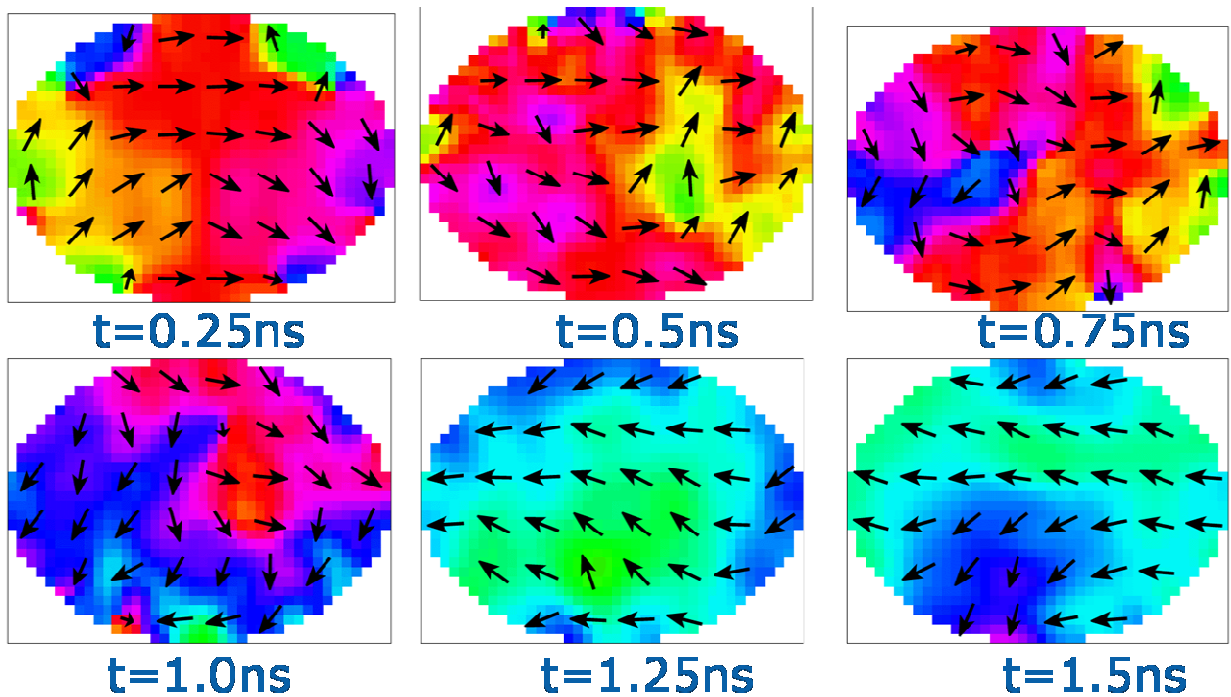

Figure 2. Snapshots of in plane magnetization distributions for the input voltage polarities A=1, B=1, C=-1. The initial state is "up" over the whole area. Electrode size 20nm, gate size 120x90nm, th=3nm, I=4mA, t=1ns.

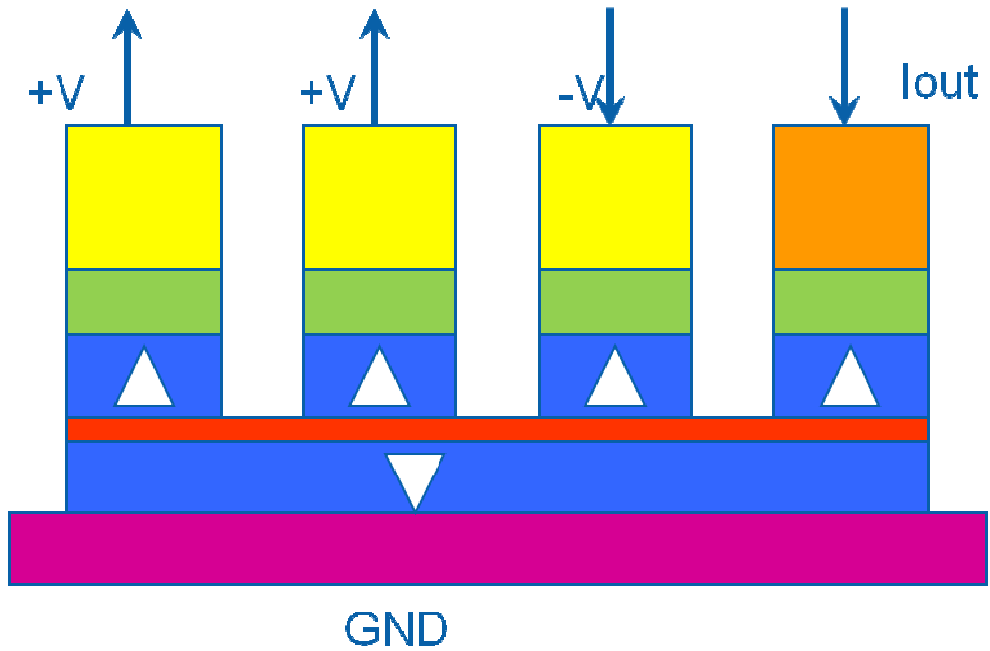

Figure 3a. Scheme of the cross-section of a STMG with perpendicular magnetization. White triangles designate the magnetization direction. The bottom layer with magnetization – the common free layer. Three input nanopillars (on the left) are designated by input voltages, and one output nanopillar (on the right) is designated by the output current. The middle layers with magnetization in nanopillars are corresponding fixed layers.

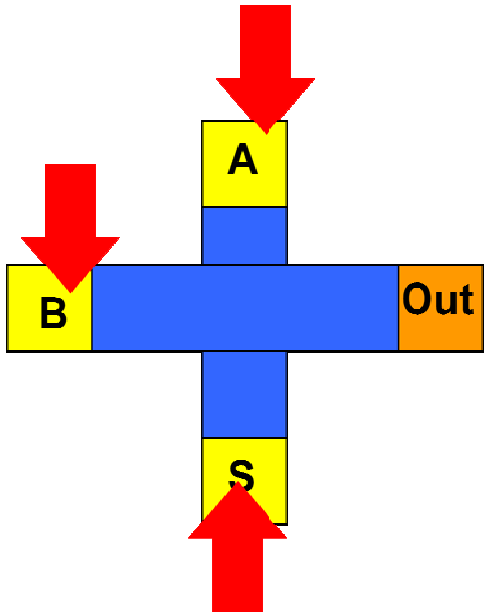

Figure 3b. Schematic of the top view of the cross STMG with perpendicular magnetization. The squares designated by "A","B", and "C" are input nanopillars. The directions of writing current are shown. The square designated by "Out" is the output nanopillar.

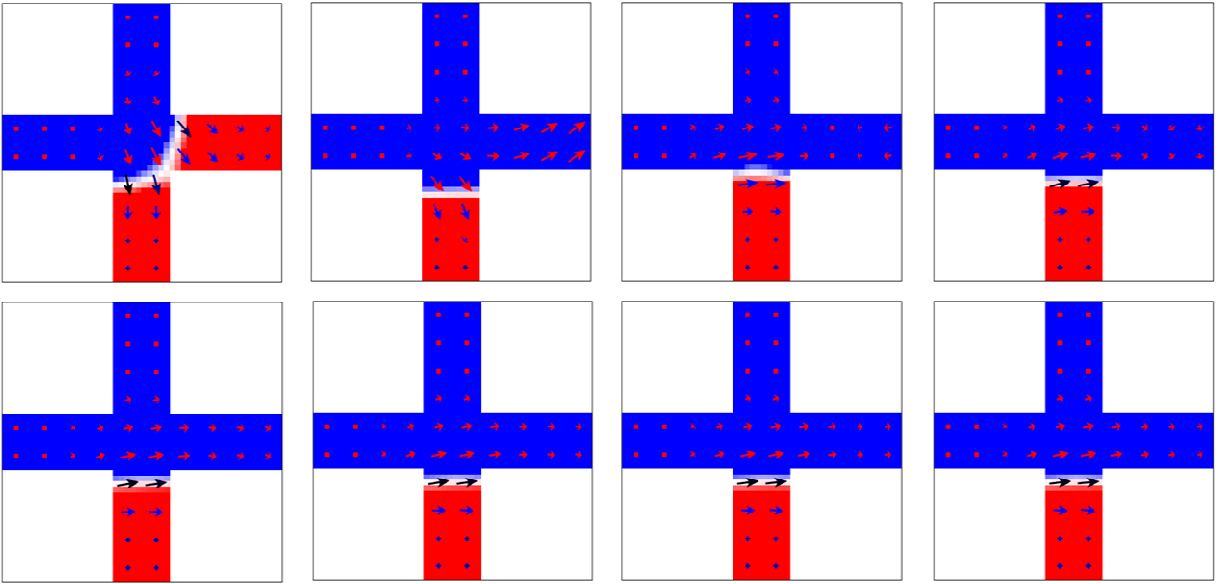

Figure 4. The distributions of magnetization in switching of a cross STMG for the input voltage polarities A=1, B=1, S=-1. The initial state is "up" over the whole area. The snapshots are done every 0.5ns. Electrode size 30nm, gate size 150x150nm, th=3nm, I=0.5mA, t=2.5ns.